# The Cancer Diaspora: Metastasis beyond the seed and soil hypothesis.

Kenneth J. Pienta[1,*], Bruce A. Robertson[2], Donald S. Coffey[1], and Russell S. Taichman[3]

[1,*] Corresponding author:
Kenneth J. Pienta
The James Buchanan Brady Urological Institute and Department of Urology,
Departments of Oncology and Pharmacology and Molecular Sciences
The Johns Hopkins School of Medicine
600 N Wolfe St, Baltimore, MD 21287
Email: kpienta1@jhmi.edu

[2]Bruce Robertson
Division of Science, Mathematics and Computing
Bard College
30 Campus Drive
Annandale-on-Hudson, New York 12504
Email: broberts@bard.edu

[1]Donald S. Coffey
The James Buchanan Brady Urological Institute and Department of Urology,
Departments of Oncology and Pharmacology and Molecular Sciences
The Johns Hopkins School of Medicine
600 N Wolfe St, Baltimore, MD 21287
Email: dcoffey@jhmi.edu

[3]Russell S. Taichman
Department of Periodontics and Oral Medicine, University of Michigan School of
Dentistry, Ann Arbor, Michigan 48109, USA.
Email: rtaich@umich.edu

**Abstract:**

Do cancer cells escape their confinement of their original habitat in the primary tumor or are they forced out by ecological changes in their home niche? Describing metastasis in terms of a simple one-way migration of cells from the primary to target organs is an insufficient concept to cover the nuances of cancer spread. A diaspora is the scattering of people away from an established homeland. To date, "diaspora" has been a uniquely human term utilized by social scientists, however, the application of the diaspora concept to metastasis may yield new biological insights as well as therapeutic paradigms. The diaspora paradigm takes into account and models several variables: the quality of the primary tumor microenvironment, the fitness of individual cancer cell migrants as well as migrant populations, the rate of bidirectional migration of cancer and host cells between cancer sites, and the quality of the target microenvironments to establish metastatic sites. Ecological scientific principles can be applied to the cancer diaspora to develop new therapeutic strategies. For example, ecological traps, habitats that lead to the extinction of a species, can be developed to attract cancer cells to a place where they can be better exposed to treatments or to cells of the immune system for improved antigen presentation. Merging the social science concept of diaspora with ecological and population sciences concepts can inform the cancer field to understand the biology of tumorigenesis and metastasis and inspire new ideas for therapy.

**Statement of Translational Relevance**

Metastases remain the major reason for the morbidity and mortality associated with cancer death. Describing metastasis as a simple migration of cells from the primary to target organs is insufficient to understand the nuances of cancer spread. To date, "diaspora", the scattering of people away from an established homeland, has been applied exclusively to describe human behavior. The application of the diaspora concept can also be applied to metastasis, which together with ecological principles may help to develop new treatment strategies. For example, ecological traps, habitats that lead to species extinction, can be developed to attract cancer cells to a place where they can be better exposed to treatments or to cells of the immune system for improved antigen presentation. Merging the social science concept of diaspora with ecological and population sciences concepts can inform the cancer field to understand the biology of tumorigenesis and metastasis and inspire new ideas for therapy.

**Metastasis viewed in terms of migration**

Do cancer cells escape their original habitat in the primary tumor or are they forced out by ecological changes in their home niche?  The latter process can be considered a diaspora and has many analogies in population and social sciences that could yield new insights into metastasis.    In 1829, Récamier recognized that cancer can spread from a primary tumor and coined the term "metastasis" from the Greek "methistemi", meaning to change or displace (Table 1) (1). Paget's "seed and soil" theory explained the non-random pattern of cancer metastasis in 1889 when he postulated that factors within the metastatic site promoted growth in the same way that fertile soil allows the successful growth of seeds (2). In a complementary hypothesis, James Ewing proposed in 1928 that cancer cells were directed to that site by the direction of lymphatic and circulatory systems (3).  From the perspective of species migration, both of these theories are correct.  Migration is subdivided into emigration (the act of leaving), migration (the act of travelling), and immigration (the process of arriving) (4). In the paradigm of migration, Ewing focused on the migration step and Paget focused on immigration (2-4). Ewing's theory accounts for the migration of prostate cancer cells to the lumbar vertebrae via Batson's plexus of draining lymph nodes and Paget's theory helps explain the organ specificity of prostate cancer metastases to bone. Fidler further refined the seed / soil hypothesis in 2003 to take into account the emigration step (5).  First, primary tumors contain heterogeneous subpopulations of cells with different angiogenic, invasive and metastatic properties (properties that promote emigration). The metastatic process is then selective for cells that can successfully survive migration to the distal target organ.  The successful proliferation of metastatic cells depends on the ability of these cells to interact and utilize the soil of the new microenvironment (5).

**Diasporas**

Describing metastasis in terms of migration alone is insufficient to cover the nuances of cancer spread (Table 1).  A diaspora, from the Greek diasporá, meaning "scattering" or "dispersion", is the migration or scattering of people away from an established homeland.  To date, "diaspora" has been applied exclusively to describe human behavior.  In ancient Greece, a diaspora referred to citizens of a conquering city-state who colonized a conquered land to assimilate the territory into the empire (6). Subsequent to the Bible's translation into Greek, "Diaspora" referred to Jewish peoples exiled from Israel by the Babylonians in 587 BCE and by the Romans from Judea in 70 CE  (7).  Since that time, diaspora has come to refer to mass-dispersions, usually involuntary, of people with common roots. Diasporas share several common features that differentiate them from migrations.  First, dispersal occurs to more than one destination (8-11). This dispersal to multiple destinations is a necessary step that eventually leads to the formation of links between the various populations of the diaspora. Second, the scattered populations must continue to retain a relationship with the homeland.  A diaspora exhibits ancestral memory – the dispersed population does not assimilate completely into the new environment, keeping a unique identity that has roots in their original home, allowing the diaspora to exist over multiple generations and exist over time. (9-12).

**Metastasis viewed as a diaspora**

Metastasis is not a simple one-way migration, but rather a dynamic dispersal of cells that are inherently linked to the primary tumor (Table 2). While migrants are usually attracted to a new country and can come from single or multiple origins, a diaspora is dispersed, usually expelled, from a single origin. At least early in tumorigenesis, metastases arise from a single site of origin, the primary tumor (13). Several types of diaspora have been described based on why the population scattered (14, 15). Emigrations have usually, but not always, been the result of harsh economic conditions or other similar hardships (4,9). Migratory diasporas may arise when individuals transit in and out of a host country but in doing so, institutions and networks become established in the hostlands, e.g. trading posts (15). Cancers appear to utilize the "trading post" diaspora analogy by forming pre-metastatic niches (16-18). Primary tumors, driven perhaps by hypoxic conditions, orchestrate the formation of pre-metastatic niches in part by the secretion of a variety of cytokines, growth factors, and exosomes to promote mobilization and recruitment of bone marrow derived cells to future metastatic sites (19-23).

The paradigm of the imperial diaspora also can be applied to metastasis. These diaspora originate as conquests, where a powerful peoples colonize (and often subjugate) an indigenous people. Metastasis is considered to be an active process of cells leaving the primary tumor site (as opposed to passive sloughing of cells into the circulation), to colonize a target organ (19-23). In an imperial diaspora, the homeland is responsible for helping to maintain an interdependent network of the diaspora. In turn, established colonies not only send resources back to the home country, but often send future generations back to the homeland for education, training, and cultural indoctrination. The exchange of resources between dispersed communities and the home country is a key aspect defining a diaspora (24-26). It has now been established in multiple experimental systems that cancer cells do traffic between tumor sites within a host. Massagué et al recently show that metastasis is a multidirectional process whereby cancer cells can seed distant sites and return to the primary tumor itself (27-29). This self-seeding, at least in preclinical models, functions to facilitate and accelerate tumor growth, angiogenesis, and recruitment of stromal cells to the tumor microenvironments. In addition, multiple host cells, including hematopoietic stem cells, mesenchymal stem cells, endothelial progenitors, cancer- associated fibroblasts, and inflammatory mononuclear cells (T-, B-, and monocytes) seem to traffic freely between tumor sites (30-35).

Another critical aspect of understanding the dynamics of diaspora development is the area that becomes the new home for the displaced population. For a displaced people, the ability and willingness for the host country to allow immigration is a key step in the formation of the diaspora. This is dependent on physical characteristics (i.e., is there room, ability for the population to thrive economically) as well as socio-political characteristics (i.e., willingness of the host community to accept the displaced peoples)(24-26). This is analogous to the favorable soil described by Paget (2, 5).

Boundary maintenance is another important criterion of a diaspora (24). Both migrant and diaspora communities maintain a collective memory of their homelands as reflected by continuing to celebrate their ethnic roots. The diaspora community, however, must maintain an identity within the host country over time. In contrast to many migrant populations, a diaspora group does not assimilate into the host population but maintains a distinct identity as so much as is possible. Cancer metastases are easily recognized in the target organ, most often observed as masses with distinct boundaries (36). This boundary maintenance blurs in migrant populations. The relationship of migrant populations with the host country tends to improve over time as the migrants assimilate into the host communities. The diaspora relationship with the host community tends to remain complicated and uneasy and better reflect the relationship of a metastatic deposit with the target organ. The metastatic deposit maintains its identity and is in constant tension with the host immune and other defense mechanisms to allow successful growth (37, 38).

**Diasporas, metastases and ecosystems**

Broadening the concept of diaspora and simultaneously applying ecological scientific principles allows a better understanding of the dispersal of cancer cells (39-41). For example, a metapopulation consists of a group of spatially separated populations of the same species that interact at some level. Metapopulation dynamics can be applied to any species in fragmented habitats; much like a diaspora from a single point of origin to multiple host countries. While cancer metastases may be thought of as metapopulations, they are actually metacommunities. Metastases are made up of multiple species of tumor and host cells interacting dynamically. Metacommunities are a network of local communities that are linked through multiple potentially interacting species or populations. Metapopulation and metacommunity dynamics allow us to study not only the growth and interactions of individual populations within different sites, but also the interactions and connectivity between sites (42).

Understanding the dispersion of metastatic cancer cells to sites throughout the body from a site of origin may use the theoretical ecology model of source sink dynamics to describe how variation in habitats (both homeland [the "source" of species] and hostland [the "sink" or habitat to which species migrates]) affect population growth or decline in metapopulations (43). Source-sink dynamics can help model how individual populations flourish or decline among different patches of habitat, e.g., a high quality habitat that on average allows the population to increase or a low quality habitat that, on its own, would not be able to support a population (referred to as a sink). Habitat quality is quite analogous to the concept of fertile or poor soil as delineated by Paget (2). Source-sink models also take into account how population numbers vary– that an excess of individuals often move from a source (primary tumor) and can continue to supply other sites (new seeds) (27-29).

The rate of by which a cancer establishes its diaspora is therefore dependent on several variables (Table 3): 1) the quality of the primary (homeland or origin) microenvironment – the more likely the microenvironment is oxygenated with a good source of nutrients, the less likely cancer cells will be driven to engage cellular programs that promote extravasation / emigration; 2) the rate of passive and active emigration.  Emigrant cancer cell populations can be divided into cells that passively slough into the circulation and cells that actively extravasate into nerves, lymphatics, or blood.  It is likely that passive emigrants may not have all of the machinery necessary to survive to successfully migrate.  Cells that have been forced to undergo an epithelial to mesenchymal transition (EMT) in response to a hypoxic environment are much more likely to survive a migration; 3) the quality and number of the target organ (hostland) microenvironments.  The quality of the soil is well documented as a critical component of successful metastasis.

Previously, it was thought that cancer cells could not differentiate between different target organs and the process of seeding was thought to be agnostic i.e., just as many cells would land in the lung as the bone depending on blood flow, etc (44).  New data suggests that this is most likely untrue on both the seed and the soil sides of the metastasis argument.  First, cancer cells can carry receptors for antigens on particular cell types, i.e, prostate cancer cells have a high level of annexin II which allows binding to osteoblasts in the endosteal niche of the bone marrow (19, 45).  Second, the elegant work of Wang and colleagues have suggested that bone marrow derived mesenchymal cells organize pre-metastatic niches (trading posts in the diaspora analogy) that provide a better soil than may actually attract cancer cells, allowing them to distinguish between high and low quality habitats (18). The source-sink model implies and models that some habitat patches may be more important to the long-term survival of the population and can be modeled by their demographic parameters or BIDE rates (birth, immigration, death, emigration) (43).

**Utilizing cancer diaspora – ecosystem concepts to target metastasis**

Viewing cancer within the context of ecology can lead to therapeutic paradigms based on network disruption across the scale of ecosystems from the cell through to the host patient (42). The diaspora concept adds to this analogy by specifically expanding this therapeutic paradigm as cancer cells metastasize and establish metacommunities. One potential therapeutic strategy would be to create ecological traps (46-48). Ecological traps are poor-quality habitats that are  highly attractive to wildlife Species select habitats based on environmental cues that typically signify a high quality habitat (46). This type of cue is analogous to signals a cancer cell receives to settle in a specific target organ. Van der Sanden and colleagues have suggested that neurotransmitters could be used to attract glioma cells to settle in a location where treatments (e.g. drugs, radiation, surgery) can be effectively or safely concentrated (near the surface, away from sensitive tissues, within surgically implanted tissue or material).  Instead of preventing metastasis, such 'attracticides' attempt to guide cancer migration   toward poor quality habitats where cancer cells can be eliminated (46).

Using traps to treat cancer is analogous to altering the ecological landscape of interactions in the body in ways in which the cells are unable to anticipate or react to and that are ultimately invisible to them.  Corollaries of this strategy include placing  familiar 'cues' in places that lead to the extinction of cancer cells, or masking attractive cues in favorable microenvironments. For glioma cells a reservoir of neurotropic chemokines could be used to attract cancer cells to an area where they could be radiated (47). For prostate cancer cells, a reservoir of SDF-1 could be temporarily inserted intravenously that attracts the cells to a one-way trap (49). Alternatively, an ecological trap could be used to expose the cancer cells to cells of the immune system, leading to increased antigen presentations (48), or disrupt the ability of metastasizing cells to recruit appropriate host cells. At present it remains a distant dream to develop methods which confine keep primary cancer cells  to their homelands  so that they find it undesirable to leave or  mask the attractiveness of the target organs (hostlands). However, the technology to create artificial tissue traps or synthetic microenvironments already does exists (48).  Another approach to killing cancer cells may be to force a diaspora from their metastatic sites where they may be protected from therapeutic intervention.  For example, Shiozawa and colleagues have demonstrated that prostate cancer cells can be mobilized out of the bone marrow microenvironment and into the unprotected circulation where they may be easier to target (45). Recognizing that genetic diversity among tumor cells to respond to different attractants is most likely, the creation of multiple evolutionary traps within a device providing multiple chemokines and/or metastatic niches should be considered (20-23, 42). The recent discovery that cancer cells continue to circulate after they metastasize suggests that this approach may offer therapeutic benefit even in patients with advanced disease (27-29).  These concepts demonstrate that merging the social science concept of diaspora with ecological and population sciences concepts can inform the cancer field to understand the biology of tumorigenesis and metastasis and inspire new ideas for therapy.

**Acknowledgements:**
This work supported by NCI U54CA143803 (KJP, DSC), CA163124, CA093900 (KJP, RST) and CA143055 (KJP).

Table 1. Examples of conceptual paradigms for metastasis

| Year | Major Authors and *concepts* | Primary Tumor | Transit of Cells | Metastases | Ref |
|---|---|---|---|---|---|
| 1829 | Récamier *First description of cancer metastases* | | | Recognized primary cancers as the source of distant masses, utilized the Greek term "methistemi", meaning "displacement" | 1 |
| 1889 | Paget *Seed and soil hypothesis* | Some "seeds" were better than others at spreading | | Seeds required a fertile "soil" to grow | 2 |
| 1928 | Ewing *Directed metastasis* | | Cells had special routes to travel | | 3 |
| 2003 | Fidler *Cells required multiple properties to escape the primary* | Cells with properties to promote escape | Cells with properties to survive the circulation | Cells with properties to flourish in a target organ | 5 |
| 2005 | Loberg and Pienta *Steps in migration* | | Transit requires emigration, migration and immigration steps | Cancer cells act as an invasive species and then naturalize and create a new microenvironment | 4 |
| 2009 | Norton and Massagué *Tumor self-seeding* | | Bidirectional flow of cells between metastases and the primary | | 27 |
| 2012 | Camacho and Pienta *Cancer and metastasis as ecosystems* | | | Metastases can be described and modeled as metacommunities | 42 |
| 2013 | Diaspora *from the Greek term "diaspora", meaning "dispersion"* | Homeland dynamics | Bidirectional migrations | Hostland Adaptation and survival | |

Table 2. A comparison of migrants, diaspora, and cancer metastases.

| Social Demography | | Cancer Demography |
|---|---|---|
| **Migrant Communities** | **Diaspora Communities** | **Cancer Metastasis** |
| Moved from a single or multiple primary homelands | Dispersed from a single homeland | Dispersed from a primary cancer |
| Attracted to new country | Pushed from homeland | Hypoxia and lack of nutrients causes pressure to leave primary |
| Host country may or may not be receptive | Host country may or may not be receptive | Target organ receptive |
| Group maintains collective memory of their homeland and culture | Group maintains collective memory of their homeland and culture | Pathologists can identify where a cancer cell originated |
| Often assimilate into the new homeland | They wish to survive as a distinct community | Metastases as distinct masses |
| Relationship with host country is complicated and uneasy but improves over time | Relationship with host country is complicated and uneasy | Immune system tries to destroy the cancer cells |
| May or may not be tied to the homeland by exchange of resources | They are tied to their homeland at many levels – exchange of resources (economic, sociopolitical) | Multiple cell type trafficking, trafficking of resources / info |

**Table 3. Development of equations may help to define the rate and success or failure of a cancer diaspora.**

| Cancer Dispersion = | Quality of Primary Microenvironment | Fitness of Migrant seeds | Quality of Metastatic sites |
|---|---|---|---|
| Cancer Diaspora = | $[O_{\Delta Q \Delta t}]$ | $[(P_{Fn\Delta t})+(A_{Fn\Delta t})]$ | $[(H1_{Qn})+(H2_{Qn})...]$ |

$O_{\Delta Q \Delta t}$ = the change in the quality ($\Delta Q$) over time ($\Delta t$) of the primary cancer site (O). Cancer cells in a highly vascularized environment with rich nutrients are presumed to be less likely to undergo an epithelial to mesenchymal transition (EMT) and leave the primary.

$P_{Fn\Delta t}$ = the number (n) and fitness (F) of passively shed cancer cells (P) over time ($\Delta t$). This represents the likelihood that a cancer cell passively shed into the circulation will survive transport to a target organ. It is likely that the fitness of a passively shed cell is less than a cell that actively exits the primary tumor through the lymphatics, nerves, or circulation.

$A_{Fn\Delta t}$ = the number (n) and fitness (F) of actively emigrant cancer cells (A) over time ($\Delta t$). This represents the likelihood that a cancer cell that actively extravasates into the circulation will survive transport to a target organ. Fitness depends on many variables, including EMT state, ability to secrete MMPs, ability to avoid anoikis, etc.

$H_{Q\,n}$ = the quality (Q) of the target organ or hostland sites (H1, H2...). Migrating cancer cells will land in multiple sites (n) within different target organs in order to immigrate. Success depends on the quality of the soil of each of these microenvironments. Prostate cancer cells, for example, are more likely to flourish in a high quality bone microenvironment then a low quality habitat such as the lung microenvironment.